\documentclass{bjour}
\usepackage{epsfig}
\newcommand {\nuc}[2]{\mbox{${}^{#1}\rm #2$}}
\begin{document}
%



\authorrunninghead{Cummins and Jones}
\titlerunninghead{Resonance Offset Tailored Pulses}




\title{Resonance Offset Tailored Pulses for NMR Quantum Computation}
\author{H.~K. Cummins}
\affil{Oxford Centre for Quantum Computation, Clarendon Laboratory, Parks
Road, Oxford OX1~3PU, UK} \email{h.cummins@physics.ox.ac.uk}
\author{J.~A. Jones\thanks{To whom correspondence should be addressed at
the Clarendon Laboratory.}} \affil{Oxford Centre for Quantum Computation,
Clarendon Laboratory, Parks Road, Oxford OX1~3PU, UK, and \\ Oxford Centre for
Molecular Sciences, New Chemistry Laboratory, South Parks Road, Oxford,
OX1~3QT, UK} \email{jonathan.jones@qubit.org}

\abstract{We describe novel composite pulse sequences which act as general
rotors and thus are suitable for nuclear magnetic resonance (NMR) quantum
computation.  The Resonance Offset Tailoring To Enhance Nutations approach
permits perfect compensation of off-resonance errors at two selected
frequencies placed symmetrically around the frequency of the RF source.}


\begin{article}
\section{Introduction}
Composite pulses \cite{Levitt:1986, Freeman:1997} play an important role in
many NMR experiments, as they allow the effects of experimental imperfections,
such as pulse length errors and off-resonance effects, to be reduced.  Such
pulses can also prove useful in NMR implementations of quantum information
processing devices, such as simple NMR quantum computers \cite{Cory:1996,
Cory:1997, Gershenfeld:1997, Jones:1998a, Jones:1998d}, where they act to
reduce systematic errors in quantum logic gates \cite{Cummins:2000a}.
Unfortunately many conventional composite pulse sequences are not appropriate
for quantum computers as they only perform well for certain initial states,
while pulse sequences designed for quantum information processing must act as
\emph{general rotors}, that is they must perform well for \emph{any} initial
state.

Composite pulses of this kind, which are sometimes called Class A composite
pulses \cite{Levitt:1986}, are rarely needed for conventional NMR experiments,
and so relatively little is known about them.  One important example is a
composite $90^\circ$ pulse developed by Tycko \cite{Levitt:1986, Tycko:1983},
which has recently been generalised to arbitrary rotation angles
\cite{Cummins:2000a}. These composite pulses give excellent compensation of
off-resonance effects at small offset frequencies, such as those found for
\nuc{1}{H} nuclei, but are of no use for the much larger off resonance
frequencies typically found for \nuc{13}{C}.

Fortunately when composite pulses are used for NMR quantum computation one
great simplification can be made: it is only necessary that the pulse sequence
perform well over a small number of discrete frequency ranges, corresponding
to the resonance frequencies of the nuclei used to implement qubits; it is
\emph{not} necessary to design pulses which work well over the whole frequency
range.  In particular it is quite common in NMR quantum computation to use at
most two spins of each nuclear species (see, for example, \cite{Marx:2000}),
and it is convenient to place the RF frequency in the centre of the spectrum,
so that the two spins have equal and opposite resonance offsets
\cite{Jones:1999a}.  Thus it suffices to tailor the composite pulse sequence
to work well at these two frequencies; the performance at all other
frequencies can be completely ignored.

Here we explain how Resonance Offset Tailoring To Enhance Nutations may be
used to produce composite pulse sequences which give perfect compensation of
off-resonance effects.  These ROTTEN pulses act as perfect general rotors at
two frequencies, offset from the RF frequency by $\pm\delta$, and are well
suited to NMR quantum computation; in combination with periods of free
precession they provide an adequate set of gates, permitting any operation to
be performed. ROTTEN pulses are simple to implement, and may be derived for
any desired resonance offset as long as $\delta \leq \sqrt{3}\nu_1$.

\section{Results}
We choose to implement our composite rotation using a sequence of three
radio-frequency (RF) pulses.  In the absence of off-resonance effects any such
pulse sequence can be described by stating two angles describing each pulse,
$\theta_j$, the nutation angle of the $j$th pulse, and $\phi_j$, the phase
angle of the nutation axis in the $xy$-plane.  In the presence of
off-resonance effects it is convenient to retain this description, except that
$\theta_j$ is now a nominal nutation angle, and the nutation axis is no longer
in the $xy$-plane (although $\phi_j$ remains a good description of the phase
angle within the plane).  It is also necessary to characterise the
off-resonance behaviour, which is conveniently parameterised using either the
off-resonance fraction $f=\delta/\nu_1$ (where $\delta$ is the off-resonance
frequency, and $\nu_1$ the nutation rate, both measured in Hertz) or,
equivalently, the RF tilt angle (the tilt angle of the nutation axis away from
the $z$-axis), given by $\tan(\Delta)=\nu_1/\delta=1/f$.  The propagator
describing a single RF pulse is then
\begin{equation}
U_j=e^{-i\theta_j(I_x\cos\phi_j+I_y\sin\phi_j+I_zf)}
\end{equation}
(where $I_x$, $I_y$ and $I_z$ and the conventional one spin product operators
\cite{Sorensen:1983}), while the overall propagator describing the three pulse
sequence is $U=U_3U_2U_1$.

In order to produce a perfectly compensated pulse it is necessary to find
values of the six angles ($\theta_1$, $\phi_1$, $\theta_2$, $\phi_2$,
$\theta_3$, and $\phi_3$) such that $U$ implements the desired rotation for
the desired value of $f$.  A general search over these six values would be a
major task, but fortunately the problem can be substantially simplified.  The
requirement that the composite pulse has \emph{identical} effects on
resonances at frequencies $\pm\delta$, so that $U(f)=U(-f)$, imposes major
restrictions on the allowed values.  Furthermore, a family of solutions exists
for which the first and last pulses are identical, thus reducing the
underlying search space to four independent values.  Examining the form of the
asymmetric response term $U(f)-U(-f)$ suggests the choices
\begin{equation}
\theta_1=\pi/\sqrt{1+f^2}
\end{equation}
and
\begin{equation}\label{eq:deltaphi}
\cos(\phi_1-\phi_2)=(1-f^2)/2.
\end{equation}
These values can then be inserted back into the expression for $U$, the result
equated with the desired propagator (neglecting any irrelevant overall phase
term), and the equations then solved for $\theta_2$ and $\phi_1$.  It is
simplest to begin by solving for a $\theta_x$ pulse, that is an ideal pulse
with nutation angle $\theta$ and phase angle $0$; pulses with any other phase
angle can then be created by simply adding a phase shift to all three pulses.
This gives the results
\begin{equation}
\theta_2=\theta/\sqrt{1+f^2}
\end{equation}
and
\begin{equation}\label{eq:phi1}
\phi_1=\pm\arccos\left(\frac{\sqrt{1+f^2}}{2}\right).
\end{equation}
(Throughout this paper we will use the positive solution of equation
\ref{eq:phi1}).  Finally combining equations \ref{eq:deltaphi} and
\ref{eq:phi1} gives
\begin{equation}\label{eq:phi2}
\phi_2=\pi-\phi_1.
\end{equation}

Examining equations \ref{eq:deltaphi} and \ref{eq:phi1} reveals a limitation
to this approach: the phase angles only have real solutions when
$|f|\leq\sqrt{3}$. It is, of course, possible that other composite pulse
families exist with larger ranges of applicability, but clearly there is a
limit beyond which any three pulse sequence will cease to function.  However,
even the limited range of $f$ values described here is far greater than
anything which can be achieved with conventional (non-tailored) composite
pulses, and is likely to be adequate for most purposes.

\begin{figure}[h]
\epsfig{file=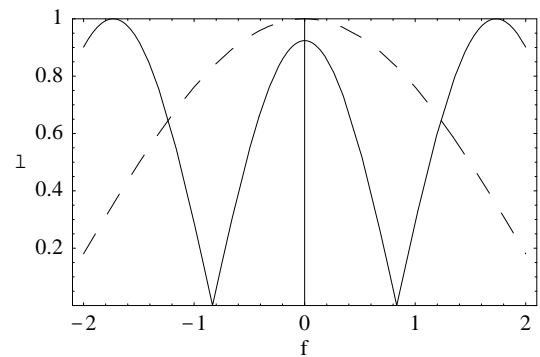,width=70mm} \caption{The fidelity, $\lambda$, of a
simple $90^\circ$ pulse (dashed line) and a ROTTEN composite $90^\circ$ pulse
for a range of values of the off-resonance fraction, $f$; the ROTTEN pulse was
tailored for the values $f=\pm\sqrt{3}$.} \label{fig:fidelity}
\end{figure}

It must, however, be remembered that ROTTEN composite pulses are only
effective at the resonance offsets for which they have been tailored; at other
resonance offsets these pulses may perform very poorly indeed.  This is shown
in figure \ref{fig:fidelity} which plots the fidelity of a ROTTEN composite
pulse sequence optimised for $f=\pm\sqrt{3}$ over a range of values of $f$. As
ROTTEN pulses are designed to act as general rotors it is necessary to use a
fidelity measure which applies over all possible starting values; we have
chosen to use the rotor fidelity, $\lambda$, defined by Levitt
\cite{Levitt:1986}.  The rotor fidelity of the simple pulse is low except at
small values of $f$, while that for the composite pulse is high in two regions
around $\pm\sqrt{3}$.  This is ideal for some implementations of NMR quantum
computation, but unlikely to prove useful in most conventional NMR
experiments.

\begin{figure*}[p]
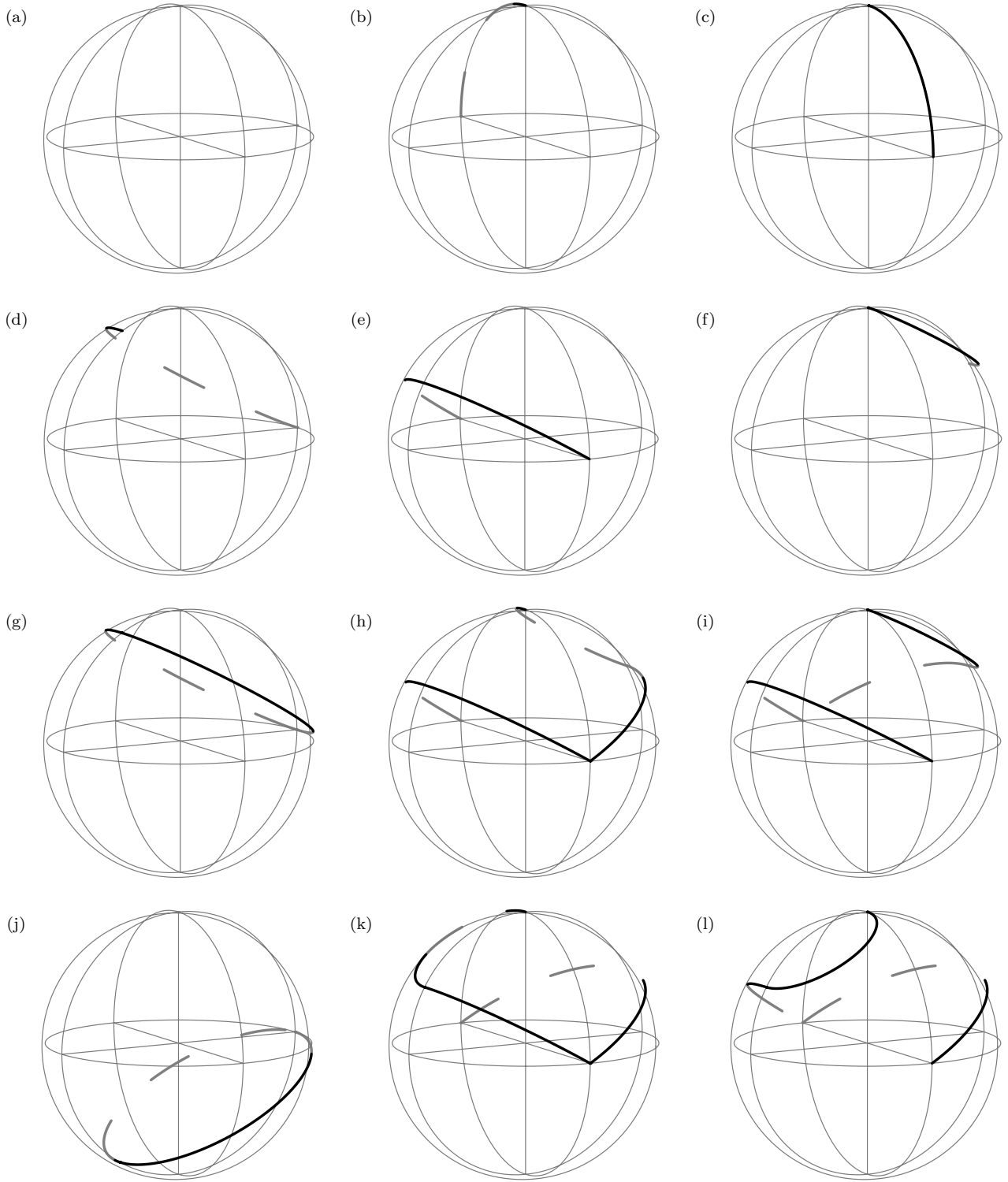

\begin{tabular}{ccc}
\raisebox{45mm}{(a)}\epsfig{file=Ix16000_ppm.eps,width=50mm}&
\raisebox{45mm}{(b)}\epsfig{file=Iy16000_ppm.eps,width=50mm}&
\raisebox{45mm}{(c)}\epsfig{file=Iz16000_ppm.eps,width=50mm}\\
\raisebox{45mm}{(d)}\epsfig{file=Ix160034_641_ppm.eps,width=50mm}&
\raisebox{45mm}{(e)}\epsfig{file=Iy160034_641_ppm.eps,width=50mm}&
\raisebox{45mm}{(f)}\epsfig{file=Iz160034_641_ppm.eps,width=50mm}\\
\raisebox{45mm}{(g)}\epsfig{file=Ix1681634_641_ppm.eps,width=50mm}&
\raisebox{45mm}{(h)}\epsfig{file=Iy1681634_641_ppm.eps,width=50mm}&
\raisebox{45mm}{(i)}\epsfig{file=Iz1681634_641_ppm.eps,width=50mm}\\
\raisebox{45mm}{(j)}\epsfig{file=Ix16816-34_641_ppm.eps,width=50mm}&
\raisebox{45mm}{(k)}\epsfig{file=Iy16816-34_641_ppm.eps,width=50mm}&
\raisebox{45mm}{(l)}\epsfig{file=Iz16816-34_641_ppm.eps,width=50mm}\\
\end{tabular}
\caption{Grapefruit plots showing magnetization trajectories for $90^\circ_x$
pulses using simple pulses and ROTTEN composite pulses optimised for an
off-resonance fraction $f=\sqrt{3}$: simple pulses (a, b, c) with no
off-resonance effects ($f=0$); simple pulses (d, e, f) in the presence of
large positive off-resonance effects ($f=\sqrt{3}$); ROTTEN pulses (g, h, i)
in the presence of large positive off-resonance effects ($f=\sqrt{3}$); ROTTEN
pulses (j, k, l) in the presence of large negative off-resonance effects
($f=-\sqrt{3}$). Initial states are $I_x$ (a, d, g, j), $I_y$ (b, e, h, k) and
$I_z$ (c, f, i, l).  Note that for a perfect $90^\circ_x$ pulse applied to
$I_x$ (a) the magnetization does not leave the $x$-axis.}
\label{fig:grapefruit}
\end{figure*}

For the remainder of this paper we will consider composite pulses tailored for
the case $f=\pm\sqrt{3}$; this is not only the limit of our approach (and so
the case where ROTTEN pulses give the greatest improvement in comparison with
conventional pulses), but also a choice which results in a particularly simple
sequence.  To achieve an ideal $\theta_\phi$ rotation the values required are
$\theta_1=\theta_3\pi/2$, $\theta_2=\theta/2$, $\phi_1=\phi_3\phi$ and
$\phi_2=\phi+\pi$. The operation of simple and ROTTEN composite $90^\circ_x$
pulses are shown by magnetization trajectories in figure \ref{fig:grapefruit}
for initial states of $I_x$, $I_y$ and $I_z$.  The magnetization trajectories
are complicated, but those for ROTTEN composite pulses terminate in the
correct locations, while simple $90^\circ_x$ pulses give extremely poor
results at such large resonance offsets.

\begin{figure*}
\begin{center}
\raisebox{45mm}{(a)}\epsfig{file=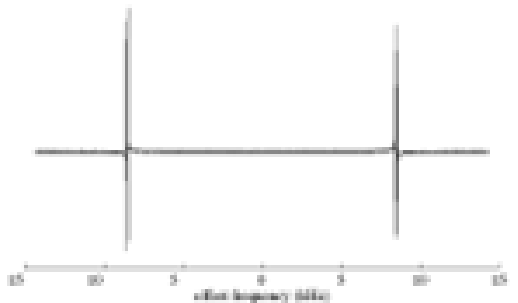,width=70mm}\qquad
\raisebox{45mm}{(b)}\epsfig{file=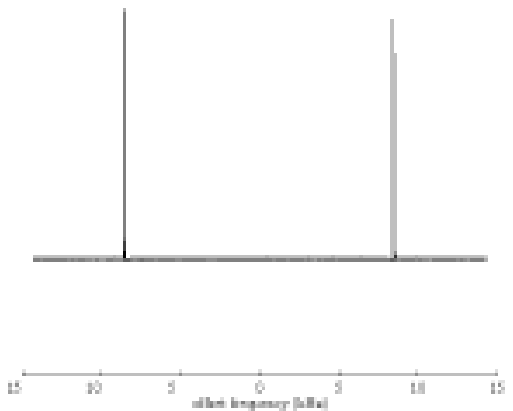,width=70mm}
\end{center}
\caption{Experimental \nuc{13}{C} spectra of \nuc{13}{C} labeled glycine in a
home built 500 MHz (\nuc{1}{H} frequency) NMR spectrometer at the Oxford
Centre for Molecular Sciences: (a) using a simple $90^\circ$ pulse; (b) using
a ROTTEN pulse.  The frequency separation between the $\rm C_\alpha$ and $\rm
C'$ multiplets was $\rm18480\,Hz$ and the RF pulse power was reduced to about
23\% of its maximum value (around $\rm13.5\,kHz$) so that the off-resonance
fraction was $f\approx\sqrt{3}$.} \label{fig:glycine}
\end{figure*}

Finally we show the performance of these pulse sequences in an actual NMR
experiment.  Figure \ref{fig:glycine} shows \nuc{13}{C} spectra of \nuc{13}{C}
labeled glycine acquired using simple and ROTTEN composite $90^\circ$
excitation pulses.  In order to emphasize the performance of our composite
pulses these spectra were acquired with the RF pulse power reduced so that the
off-resonance fraction was $\sqrt{3}$.  Under these circumstances a simple
$90^\circ$ excitation results in phase errors of $\pm90^\circ$, while the
ROTTEN pulse should give perfect compensation.  In practice small phase errors
are observed in the ROTTEN spectra; calculations suggest that these arise from
pulse length errors.

\section{Conclusions}
Resonance offset tailoring provides a simple and effective approach for
removing large off-resonance effects in systems where NMR resonances occur at
two well separated frequencies.  Unlike most conventional composite pulses,
ROTTEN pulses provide theoretically perfect compensation, and act as perfect
rotors.  These properties make them well suited to NMR quantum computation.
They are, however, unlikely to be useful in more conventional NMR experiments.

\begin{acknowledgment}
We thank M.~Bowdrey and A.~Pittenger for helpful discussions.  H.K.C. thanks
NSERC (Canada) and the TMR programme (EU) for their financial assistance.
J.A.J. is a Royal Society University Research Fellow.  This is a contribution
from the Oxford Centre for Molecular Sciences, which is supported by the UK
EPSRC, BBSRC, and MRC.
\end{acknowledgment}

\end{article}
\end{document}